\documentclass[10pt,twocolumn]{IEEEtran}
\IEEEoverridecommandlockouts
\IEEEoverridecommandlockouts
\usepackage{cite}
\usepackage{amsmath,amssymb,amsfonts}
\usepackage{algorithmic}
\usepackage{soul}
\usepackage[ruled,vlined]{algorithm2e}
\usepackage{psfrag}
\usepackage[table]{xcolor}
\usepackage{color}
\usepackage{graphicx}
\usepackage{acronym}
\usepackage{xcolor}
\usepackage{comment}
\usepackage{longtable}
\usepackage{graphicx}
\usepackage{booktabs, tabularx}
\usepackage{tikz}
\usepackage{pgfplots}
\usepackage{bbm}

\definecolor{darkred}{RGB}{128, 0, 34}

\begin{document}

\title{Model Proficiency in Centralized \\Multi-Agent Systems: A Performance Study}

\author{Anna Guerra,~\IEEEmembership{Member,~IEEE,} Francesco Guidi,~\IEEEmembership{Member,~IEEE,} Pau Closas,~\IEEEmembership{Senior Member,~IEEE,} \\Davide Dardari,~\IEEEmembership{Fellow,~IEEE,} and Petar M. Djuri\'c,~\IEEEmembership{Life Fellow,~IEEE}
\thanks{Anna Guerra and Davide Dardari are with the University of Bologna - DEI ``Guglielmo Marconi", Italy. E-mail: \{anna.guerra3, davide.dardari\}@unibo.it \\ Francesco Guidi is with the CNR, IEIIT, Bologna, Italy. E-mail: francesco.guidi@cnr.it. Pau Closas is with ECE, Northeastern University, Boston, USA. E-mail: closas@northeastern.edu. Petar M.~Djuri\'c is with ECE, Stony Brook University, Stony Brook, NY 11794, USA.
E-mail: petar.djuric@stonybrook.edu.  This work was partially supported by the European Union under the Italian National Recovery and Resilience Plan (NRRP) of NextGenerationEU, partnership (Mission 4  – Component 2  -Investment 1.1) Prin 2022 (No. 104, 2/2/2022, CUP J53C24002790006), under ERC Grant no. 101116257 (project CUE-GO: Contextual Radio Cues for Enhancing Decision Making in Networks of autonomous agents), and by the NSF under Awards 2212506, 1845833, 2326559 and 2530870.}}%


\maketitle

\begin{abstract}
Autonomous agents are increasingly deployed in dynamic environments where their ability to perform a given task depends on both individual and team-level proficiency. While \ac{PSA} has been studied for single agents, its extension to a team of agents remains underexplored. This letter addresses this gap by presenting a framework for team \ac{PSA} in centralized settings. We investigate three metrics for centralized team PSA: the  \ac{MPB}, the \ac{KS} statistic, and the \ac{KL} divergence. These metrics quantify the discrepancy between predicted and actual measurements. We use the KL divergence as a reference metric since it compares the true and predictive distributions, whereas the \ac{MPB} and \ac{KS} provide efficient indicators for in situ assessment. Simulation results in a target tracking scenario demonstrate that both \ac{MPB} and \ac{KS} metrics accurately capture model mismatches, align with the \ac{KL} divergence reference, and enable real-time proficiency assessment.
\end{abstract}

\acresetall
\bstctlcite{IEEEexample:BSTcontrol}

\newcommand{\iModel} {m}
\newcommand{\iTime} {t}
\newcommand{\iTimePrev} {\iTime-1}
\newcommand{\iRangeott} {1:\iTime}
\newcommand{\iRangeotT} {1:\T}
\newcommand{\iRangeottprev} {1:\iTimePrev}
\newcommand{\iTimeCondttprev} {\iTime \lvert \iTimePrev}
\newcommand{\iTimeCondtprevtprev} {\iTimePrev \lvert \iTimePrev}
\newcommand{\iTimeCondtt} {\iTime \lvert \iTime}
\newcommand{\iAgent} {i}
\newcommand{\jAgent} {j}
\newcommand{\iMAgent} {i'}
\newcommand{\iSampleK} {k}
\newcommand{\iSampleL} {\ell}
\newcommand{\iData} {j}
\newcommand{\iMeas} {q}
\newcommand{\iMC} {n}
\newcommand{\iPSA}{\mathrm{MPB}}
\newcommand{\iKS}{\mathrm{KS}}
\newcommand{\iKSPSA}{\mathrm{KS-MPB}}
\newcommand{\iKL}{\mathrm{KL}}

\newcommand{\NSampleK}{N_\text{K}}
\newcommand{\NSampleL}{N_\text{L}}
\newcommand{\NModel} {N_\text{M}}
\newcommand{\NMeas} {N_\text{O}}
\newcommand{\NMeastot} {N_{\text{D}}}
\newcommand{\NAgents} {N_{\text{A}}}
\newcommand{\NNAgents} {N^{(\iAgent)}}
\newcommand{\NState} {N_{\text{S}}}
\newcommand{\NMC}{N_{\text{MC}}}
\newcommand{\T}{N_{\text{hor}}}

\newcommand{\pdf} {f}
\newcommand{\cdf} {F}
\newcommand{\approxpdf} {\tilde{\pdf}}
\newcommand{\approxcdf} {\tilde{\cdf}}

\newcommand{\Hnull}{\mathcal{H}_0}
\newcommand{\Ha}{\mathcal{H}_1}

\newcommand{\tra} {T}
\newcommand{\expect} {\mathbb{E}}
\newcommand{\trace} {\mathsf{tr}}
\newcommand{\determ} {\mathsf{det}}
\newcommand{\diag} {\operatorname{diag}}
\newcommand{\erf}{\operatorname{erf}}
\newcommand{\argmin}{\operatorname{argmin}}

\newcommand{\Normal} {\mathcal{N}}
\newcommand{\Uniform} {\mathcal{U}}

\newcommand{\Model} {\mathcal{M}}
\newcommand{\CModel} {\mathcal{M}_0}
\newcommand{\Mnull} {\mathcal{M}_0}
\newcommand{\Mone} {\mathcal{M}_1}
\newcommand{\Mtwo} {\mathcal{M}_2}
\newcommand{\bthetanull} {\boldsymbol{\theta}_0}
\newcommand{\bthetaone} {\boldsymbol{\theta}_1}
\newcommand{\bthetatwo} {\boldsymbol{\theta}_2}





\newcommand{\btheta} {\boldsymbol{\theta}}
\newcommand{\RVx} {\mathsf{x}} 
\newcommand{\RVbx} {\boldsymbol{\mathsf{x}}} 
\newcommand{\RVhbx} {\hat{\boldsymbol{\mathsf{x}}}} 
\newcommand{\RVbX} {\boldsymbol{\mathsf{X}}}
\newcommand{\RVzt} {\mathsf{z}_{\iTime}}
\newcommand{\RVy} {\mathsf{y}}
\newcommand{\RVhy} {\hat{\mathsf{y}}}
\newcommand{\RVhyt} {\RVhy_{\iTime}}
\newcommand{\RVby} {\boldsymbol{\mathsf{y}}}
\newcommand{\RVhby} {\hat{\boldsymbol{\mathsf{y}}}}
\newcommand{\RVbY} {\boldsymbol{\mathsf{Y}}}
\newcommand{\RVe}{\mathsf{e}}
\newcommand{\RVbe}{\boldsymbol{\mathsf{e}}}
\newcommand{\RVu}{\mathsf{u}}
\newcommand{\RVbu}{\boldsymbol{\mathsf{u}}}
\newcommand{\RVv}{\mathsf{v}}
\newcommand{\RVbv}{\boldsymbol{\mathsf{v}}}
\newcommand{\RVpos}{\boldsymbol{\mathsf{p}}}
\newcommand{\RVspeed}{\dot{\RVpos}}

\newcommand{\x} {x}
\newcommand{\bx} {\boldsymbol{x}}
\newcommand{\bX} {\boldsymbol{X}}
\newcommand{\y} {y}
\newcommand{\hy} {\hat{y}}
\newcommand{\by} {\boldsymbol{y}}
\newcommand{\bhy} {\hat{\boldsymbol{y}}}
\newcommand{\bY} {\boldsymbol{Y}}
\newcommand{\kgain}{\kappa}
\newcommand{\bkgain}{\boldsymbol{\kappa}}
\newcommand{\vary}{s^2}
\newcommand{\bvary}{\boldsymbol{S}}
\newcommand{\e}{e}
\newcommand{\be}{\boldsymbol{e}}
\newcommand{\bu}{\boldsymbol{u}}
\newcommand{\bv}{\boldsymbol{v}}
\newcommand{\spos}{p}
\newcommand{\pos}{\boldsymbol{p}}
\newcommand{\speed}{\dot{\pos}}
\newcommand{\bA}{\boldsymbol{A}}
\newcommand{\bQ}{\boldsymbol{Q}}
\newcommand{\bQQ}{\bar{\bQ}}
\newcommand{\bH}{\boldsymbol{H}}
\newcommand{\bhit} {\boldsymbol{h}_{\iAgent,\iTime}}
\newcommand{\RVbhit} {\boldsymbol{\mathsf{h}}_{\iAgent,\iTime}}
\newcommand{\bhjt} {\boldsymbol{h}_{\jAgent,\iTime}}
\newcommand{\bht} {\boldsymbol{h}_{\iTime}}
\newcommand{\hit} {h_{\iAgent,\iTime}}
\newcommand{\RVhit} {\mathsf{h}_{\iAgent,\iTime}}
\newcommand{\bRi} {\boldsymbol{R}^{(\iAgent)}}
\newcommand{\bR} {\boldsymbol{R}}
\newcommand{\m} {m}
\newcommand{\bm} {\boldsymbol{m}}

\newcommand{\bP} {\boldsymbol{P}}

\newcommand{\dt} {T_{\text{s}}}

\newcommand{\RVbxt} {\RVbx_{\iTime}}
\newcommand{\RVhbxt} {\RVhbx_{\iTime}}
\newcommand{\RVbxtprev} {\RVbx_{\iTimePrev}}
\newcommand{\RVxt} {\RVx_{\iTime}}
\newcommand{\RVxtprev} {\RVx_{\iTimePrev}}
\newcommand{\bxt} {\bx_{\iTime}}
\newcommand{\bxtprev} {\bx_{\iTimePrev}}
\newcommand{\xt} {\x_{\iTime}}
\newcommand{\xtprev} {\x_{\iTimePrev}}
\newcommand{\bxtpred} {\bx_{\iTimeCondttprev}}
\newcommand{\bxtupd} {\bx_{\iTimeCondtt}}
\newcommand{\mtpred} { \m_{\iTimeCondttprev}} 
\newcommand{\mtupd} { \m_{\iTimeCondtt}} 
\newcommand{\mtprevupd} { \m_{\iTimeCondtprevtprev}} 
\newcommand{\bmtpred} {\bm_{\iTimeCondttprev}} 
\newcommand{\bmtpredi} {\bm_{\iAgent;\iTimeCondttprev}} 
\newcommand{\vartpred} { \sigma^2_{\iTimeCondttprev}} 
\newcommand{\vartupd} { \sigma^2_{\iTimeCondtt}} 
\newcommand{\vartprevupd} {\sigma^2_{ \iTimeCondtprevtprev}} 
\newcommand{\bvartpred} {\bP_{\iTimeCondttprev}} 
\newcommand{\bvartpredi} {\bP_{\iAgent;\iTimeCondttprev}} 

\newcommand{\RVxo} {\RVx_{0}}
\newcommand{\RVbxo} {\RVbx_{0}}
\newcommand{\xo} {\x_{0}}
\newcommand{\bxo} {\bx_{0}}

\newcommand{\RVyt} {\RVy_{\iTime}} 
\newcommand{\RVyotprev}{\RVy_{\iRangeottprev}} 
\newcommand{\RVbyotprev}{\RVby_{\iRangeottprev}} 
\newcommand{\RVyit} {\RVy_{\iAgent,\iTime}} 
\newcommand{\RVyjt} {\RVy_{\jAgent,\iTime}} 
\newcommand{\RVbyt} {\RVby_{\iTime}} 
\newcommand{\RVbyit} {\RVby_{\iAgent,\iTime}} 
\newcommand{\RVbyjt} {\RVby_{\jAgent,\iTime}} 
\newcommand{\RVhbyt} {\RVhby_{\iTime}} 
\newcommand{\yt} {\y_{\iTime}} 
\newcommand{\byt} {\by_{\iTime}}
\newcommand{\byit} {\by_{\iAgent,\iTime}}
\newcommand{\hyt} {\hat{y}_{\iTime}} 
\newcommand{\hbyt} {\bhy_{\iTime}}
\newcommand{\yot} {\y_{\iRangeott}} 
\newcommand{\yotprev} {\y_{\iRangeottprev}} 
\newcommand{\byotprev} {\by_{\iRangeottprev}} 
\newcommand{\byotprevi} {\by_{\iRangeottprev}^{(\iAgent)}} 
\newcommand{\ytpred} {\y_{\iTimeCondttprev}} 
\newcommand{\bytpred} {\boldsymbol{y}_{\iTimeCondttprev}} 

\newcommand{\kgaint} {\kgain_{\iTime}}
\newcommand{\bkgaint} {\bkgain_{\iTime}}
\newcommand{\varytpred} {\vary_{\iTimeCondttprev}}
\newcommand{\bvarytpred} {\bvary_{\iTimeCondttprev}}
\newcommand{\bvarytpredi} {\bvary_{\iTimeCondttprev}^{(\iAgent)}}
\newcommand{\RVbet} {\RVbe_{\iTime}}
\newcommand{\et} {\e_{\iTime}}
\newcommand{\bet} {\be_{\iTime}}
\newcommand{\SEt} {\text{SE}_{\iTime}}
\newcommand{\Jac} {\boldsymbol{J}}

\newcommand{\RVut} {\RVu_{\iTime}}
\newcommand{\RVbut} {\RVbu_{\iTime}}
\newcommand{\ut} {u_{\iTime}}
\newcommand{\but} {\bu_{\iTime}}
\newcommand{\RVvt} {\RVv_{\iTime}}
\newcommand{\RVbvt} {\RVbv_{\iTime}}
\newcommand{\RVbvit} {\RVbv_{\iAgent,\iTime}}
\newcommand{\RVbvjt} {\RVbv_{\jAgent,\iTime}}
\newcommand{\RVvit} {\RVv_{\iAgent,\iTime}}
\newcommand{\RVvjt} {\RVv_{\jAgent,\iTime}}
\newcommand{\vt} {v_{\iTime}}
\newcommand{\bvt} {\boldsymbol{v}_{\iTime}}
\newcommand{\varu} {\sigma^2_u}
\newcommand{\varv} {\sigma^2_v}
\newcommand{\varvit} {\sigma^2_{v, \iAgent, \iTime}}
\newcommand{\varvjt} {\sigma^2_{v, \jAgent, \iTime}}

\newcommand{\RVpost}{\RVpos_{\iTime}}
\newcommand{\RVspeedt}{\RVspeed_{\iTime}}

\newcommand{\PSA}{\mathcal{P}}
\newcommand{\bPSA}{\boldsymbol{\PSA}}
\newcommand{\PSAt}{\PSA_{\iTime}}
\newcommand{\PSATt}{\PSA_{\mathsf{T},\iTime}}
\newcommand{\bPSAt}{\bPSA_{\iTime}}
\newcommand{\dPSAt}{d_{\iPSA, \iTime}}
\newcommand{\ndPSAt}{\bar{d}_{\iPSA, \iTime}}

\newcommand{\RVdKS}{\mathsf{d}_{\iKS}}
\newcommand{\dKS}{d_{\iKS}}
\newcommand{\RVdKSt}{\mathsf{d}_{\iKS, \iTime}}
\newcommand{\RVadKSt}{\bar{\mathsf{d}}_{\iKS, \iTime}}
\newcommand{\dKSt}{d_{\iKS, \iTime}}
\newcommand{\dKSott}{d_{\iKS, \iRangeott}}
\newcommand{\dKSotT}{d_{\iKS, \iRangeotT}}

\newcommand{\dKSPSAt}{d_{\iKSPSA,\iTime}}

\newcommand{\dKL}{d_\iKL}
\newcommand{\dKLt}{d_{\iKL, \iTime}}
\newcommand{\dKLti}{d_{\iKL, \iTime}^{(\iAgent)}}


\newcommand{\posit}{\pos_{\iAgent, \iTime}}
\newcommand{\posi}{\pos_{\iAgent}}
\newcommand{\posj}{\pos_{\jAgent}}
\newcommand{\post}{\pos_{\iTime}}
\newcommand{\hpost}{\hat{\pos}_{\iTime}}
\newcommand{\speedt}{\speed_{\iTime}}

\acrodef{AA}{autonomous agent}
\acrodef{AI}{artificial intelligence}
\acrodef{AOA}{angle-of-arrival}
\acrodef{A2A}{agent-to-agent}
\acrodef{COA}{curvature-of-arrival}
\acrodef{CSI}{channel state information}
\acrodef{CRLB}{Cram\'er-Rao Lower Bound}
\acrodef{CDF}{cumulative density function}
\acrodef{DRN}{dynamic radar network}
\acrodef{EKF}{extended Kalman filter}
\acrodef{EA}{evolutionary algorithm}
\acrodef{ECDF}{empirical cumulative density function}
\acrodef{GA}{genetic algorithm}
\acrodef{KF}{Kalman filter}
\acrodef{KL}{Kullback-Leibler}
\acrodef{KS}{Kolmogorov-Smirnov}
\acrodef{LS}{least-squares}
\acrodef{LOS}{line-of-sight}
\acrodef{MPB}{measurement prediction bound}
\acrodef{MDP}{Markov decision process}
\acrodef{MLE}{maximum likelihood estimator}
\acrodef{MAP}{maximum {\em a posteriori}}
\acrodef{MMSE}{minimum mean square error}
\acrodef{MSE}{mean square error}
\acrodef{NLOS}{non line of sight}
\acrodef{PF}{particle filter}
\acrodef{PSGD}{projected steepest gradient descent}
\acrodef{PDF}{probability density function}
\acrodef{PSO}{particle swarm optimization}
\acrodef{PSA}{proficiency self-assessment}
\acrodef{RSSI}{received signal strength indicator}
\acrodef{RMSE}{root mean square error}
\acrodef{RV}{random variable}
\acrodef{SAR}{synthetic aperture radar}
\acrodef{SSM}{State-Space Model}
\acrodef{SLAM}{simultaneous localization and mapping}
\acrodef{SNR}{signal-to-noise ratio}
\acrodef{UAV}{unmanned aerial vehicle}
\acrodef{UKF}{unscented Kalman filter}
\acrodef{UPF}{unscented particle filter}

\vspace{-0.2cm}
\section{Introduction}
\vspace{-0.1cm}
 
 In recent years, \acp{AA} have rapidly expanded as a means to address increasingly complex problems under real-time constraints. To fully harness their potential, there is a need to design coordinated teams of \acp{AA} that can carry out demanding tasks while evaluating both their individual and team proficiency, thus engaging in ongoing learning throughout their operation \cite{GueEtAl:J22,zhang2020self,GueGuiDarDju:J23}.

\Ac{PSA} is not a new concept and has already been applied in multiple disciplines, such as explainable \ac{AI} \cite{mohseni2021multidisciplinary}, safety applications \cite{dearden2004real}, and robotics \cite{triebel2016driven, frasca2020can}. Several recent surveys provide a comprehensive overview of this evolving field \cite{ConEtAl:J23, cao2023robot, norton2022metrics}. Broadly speaking, \ac{PSA} refers to an agent's ability to anticipate, estimate, and evaluate its own capacity to accomplish a task \cite{norton2022metrics}. This ability enables \acp{AA} to refine their decision-making and interactions while operating in dynamic, time-varying environments. In this work, \ac{PSA} validates the reliability of the models that agents use to interpret their surroundings \cite{bernardo2009bayesian}. 

\ac{PSA} methodologies have traditionally been divided into three main categories, according to whether they rely on an expert (e.g., human), machine learning, or statistics \cite{ConEtAl:J23}. This work focuses on test-based PSA performed \emph{in situ} (during execution), which leverages statistical techniques \cite{guyer2022will,israelsen2020machine,fitzgerald2019human,fleming2017self}. In signal processing, model validation frequently relies on statistical hypothesis testing to assess regression and generalized linear models \cite{bernardo2009bayesian,closas2009assessing,djuric2009model,djuric2010assessment}, as well as to identify regime-switching models \cite{li2023differentiable, el2021particle}. For example, \cite{djuric2010assessment} proposed a method for evaluating dynamic nonlinear models based on empirical and predictive cumulative distributions together with the \ac{KS} statistic. Similarly, in \cite{DjuClo:C19}, Bayesian \ac{CRLB} derived from predictive distributions was employed to quantify model proficiency.
Overall, \ac{PSA} is fundamental to improving efficiency, adaptability, resource allocation, fault tolerance, and system reliability. However, while significant progress has been made in equipping individual agents with \ac{PSA} capabilities, advancing team-level proficiency assessment remains underexplored.

This letter presents an analysis of team proficiency in centralized network architectures, combining statistical and information-theoretic metrics. 
Specifically, we examine the application of the \ac{MPB}, the \ac{KS} statistic, and the \ac{KL} divergence to a team of \acp{AA}. 
Their effectiveness is evaluated in a moving-target tracking scenario, demonstrating that overall tracking performance can be enhanced by team proficiency, which identifies agents that rely on inaccurate observation models.

\vspace{-0.2cm}
\section{Research Problem}
\vspace{-0.1cm}

Let us consider a network of $\NAgents$ \acp{AA}, where the agents observe a hidden (unknown) state, denoted with the \ac{RV} $\RVbxt$ at time instant $\iTime$. Each agent is assumed to collect measurements of the environment's state, which are then shared across the network. 
The joint observation vector is given by $\RVbyt=\left[\RVby_{1,\iTime}^\tra,\,\cdots,\, \RVby_{\iAgent,\iTime}^\tra,\,\cdots,\,\RVby_{\NAgents,\iTime}^\tra \right]^\tra,$ with $\RVbyit$ being the observations collected by the $\iAgent$-th sensors. By processing the information gathered by the agents, the network aims to estimate the hidden state $\RVhbxt$. Formally, this estimate is given by the conditional expectation of the state $\RVhbxt = \mathbb{E}\left[ \RVbxt \lvert \RVbyotprev; \Model\right]$, where the expectation is taken in accordance with the \ac{SSM} defined as
\begin{align}
\Model &\triangleq \left\{\Model_x,\, \Model_y \right\}\nonumber \\
&=\left\{\pdf_{\RVbxt \lvert  \RVbxtprev}(\bxt \lvert  \bxtprev,\, \btheta_x), \,  \pdf_{\RVbyit \lvert \RVbxt}(\byit \lvert \bxt,\, \btheta_y) \right\}\,,
\end{align}
with $\Model_x$, $\Model_y$ being models of the state and observation dynamics, respectively, $\btheta=\left\{\btheta_x,\, \btheta_y \right\}$ being a vector of parameters, $\pdf_{\RVbxt \lvert  \RVbxtprev}(\cdot)$ and $\pdf_{\RVbyit \lvert \RVbxt}(\cdot)$ being \acp{PDF} that describe the state evolution and observations. 
 
The standard way to assess estimation accuracy is by comparing the estimate $\RVhbxt$ with the ground truth state $\RVbxt$, which is usually not available. Consequently, performance is often assessed indirectly through the model's ability to predict future measurements $\RVbyt$ from past observations $\RVbyotprev$.
The predicted observation at time $\iTime$ is $\RVhbyt = g_{\Model}(\RVbyotprev) =\mathbb{E}\left[ \RVbyt \lvert \RVbyotprev; \Model\right]$. 

To this end, each agent uses a set of $\NModel$ statistical models stored in a library denoted with $\mathcal{L}=\left\{\CModel, \Model_1,\, \Model_2,\, \ldots,\, \Model_{\iModel}, \ldots, \, \Model_{\NModel-1} \right\}$ with the true generative model, $\CModel$, being included in $\mathcal{L}$ \cite{bernardo2009bayesian}. 

This paper addresses the fundamental question: how can a centralized team of \acp{AA} assess its team proficiency in predicting observations using a metric that quantifies the discrepancy between predicted and actual measurements without relying on knowledge of the true hidden state, and that guides the selection of the most appropriate model?

\section{Proficiency Analysis}
To address the previous question, we start by introducing three different metrics for proficiency assessment. We first discuss the \ac{MPB}, originally proposed in \cite{DjuClo:C19} for a single agent. This metric is inspired by the Bayesian \ac{CRLB} and represents a fundamental limit for the measurement error when predicting a new observation under a given model. Then, we recall the \ac{KS} statistic, which compares the \ac{ECDF} of the predictive distribution with that of the actual measurement distribution \cite{closas2009assessing}. Unlike the previous approach, which yields a score, proficiency here is assessed through a hypothesis test (significance testing).
Finally, we consider the \ac{KL} divergence that directly compares the \ac{PDF} of the prediction with the true \ac{PDF} of the measurement. 

\paragraph{Measurement Prediction Bound (MPB)}
Following the definition of \ac{PSA} presented in \cite{DjuClo:C19} for scalar observations, we define the {\em in situ} \ac{MPB} as in \cite[Eq. 6]{DjuClo:C19}
\begin{align}
\! \! \! \PSAt\left( \Model | \yotprev \right) &= - \expect\left[ \frac{\partial ^2\, \ln \pdf_{\RVyt | \RVyotprev}\left( \yt | \yotprev;\,\Model \right)}{\partial \, \yt^2} \right],
\label{eq:proficiency_insitu} 
\end{align}
where the expectation $\expect \left[\cdot \right]$ is taken with respect to the predictive density $\pdf_{\RVyt\, \mid \RVyotprev}\left(\y_t \mid \yotprev; \Model \right)$. 
Under standard regularity conditions (differentiability, integrability, vanishing boundary terms) and for any predictor $\RVhyt$ with a finite second moment, the Bayesian information inequality (Van Trees) implies
\cite{gill1995applications,van2007bayesian}
\begin{align}\label{eq:PSAMSE}
\bigg(\,\mathbb{E}\!\left[( \RVhyt - \RVyt )^{2}\right]\,\bigg)^{-1}
\;\le\; \PSAt\!\left(\Model \mid \yotprev \right),
\end{align}
i.e., the \ac{MPB} upper-bounds the inverse MSE of the measurement prediction.

Extending the {\em in situ} \ac{MPB} of \eqref{eq:proficiency_insitu} to the joint measurement vector related to the team, we have 
\begin{align}\label{eq:proficiency_insitu_vector}
\bPSAt\left( \Model \lvert \byotprev \right) = &\expect\left[  \nabla_{ \RVbyt } \ln \pdf_{\RVbyt | \RVbyotprev}\left( \byt | \byotprev; \Model \right) \right. \nonumber\\
&\left.\times \nabla_{\RVbyt} \ln \pdf_{\RVbyt | \RVbyotprev}\left( \byt | \byotprev; \Model \right)^\tra \right], 
\end{align}
with $\nabla_{\RVbyt}=\frac{\partial}{\partial \byt}= \left[\frac{\partial}{\partial \y_{1, t}},\, \frac{\partial}{\partial \y_{2, t}}, \ldots, \, \frac{\partial}{\partial \y_{N_{\text{D}}, t}} \right]^\tra$ and $N_{\text{D}}$ being the size of the observation vector. 
Then, for any predictor with a finite second moment,
\begin{align}
\label{eq:vector_bound}
\mathbb{E}\!\big[(\widehat{\RVby}_t-\RVbyt)(\widehat{\RVby}_t-\RVbyt)^{\!\top}\big]
~\succeq~ \bPSAt\!\left(\Model \mid \byotprev \right)^{-1},
\end{align}
i.e., the vector \ac{MPB} is an information matrix whose inverse lower-bounds the prediction-error covariance.

To represent the \ac{MPB} in a scalar way, one can take the trace of \eqref{eq:proficiency_insitu_vector} as
\begin{align}\label{eq:prof_team}
\PSATt\left( \Model \lvert \byotprev \right)=\frac{1}{\NAgents}\trace\left(\bPSAt\left( \Model \lvert \byotprev \right)\right) \, ,
\end{align} 
where the normalization in \eqref{eq:prof_team} allows for the computation of the average \ac{MPB} of the team. The individual \acp{MPB} of single agents is given by the diagonal elements of the matrix $\bPSAt\left( \Model \lvert \byotprev \right)$ in \eqref{eq:proficiency_insitu_vector}.

When there is a model mismatch between the model used for data generation and the one used for estimation, the \ac{MPB} alone is not sufficient to assess an agent's proficiency. Therefore,  we measure proficiency via the normalized absolute deviation 
\begin{align}\label{eq:normdistance}
    \ndPSAt\left(\Model \right) 
    &= \frac{\lvert \PSAt^{-1}\left( {\Model|\RVyotprev} \right) -{e}_t^2\left(\Model \right) \rvert}{\lvert \PSAt^{-1}\left( {\Model| \RVyotprev } \right) \rvert}\,,
\end{align}
where $e_t^2(\Model)=\lvert \yt-\hyt(\Model)\rvert^2$. This dimensionless score compares the inverse MPB (an MSE lower bound) with the observed squared prediction error; smaller values indicate closer agreement.

\begin{algorithm}[t!]
\SetAlgoLined
Initialize  $\boldsymbol{m}_{0 \lvert 0}, \boldsymbol{P}_{0 \lvert 0}, \forall \iAgent \in \mathcal{N}$; \\
  \For{$\iTime=1, \ldots, \T$}{
  $\boldsymbol{m}_{\iTime \lvert \iTime-1} = \boldsymbol{A}\, \boldsymbol{m}_{\iTime-1 \lvert \iTime-1}$; \\
$\boldsymbol{P}_{ \iTime \lvert \iTime-1} = \boldsymbol{A}\, \boldsymbol{P}_{\iTime-1 \lvert \iTime-1}\, \boldsymbol{A}^\tra + \boldsymbol{Q}\,$;\\
$\boldsymbol{e}_{\iTime}=\boldsymbol{y}_{\iTime}-\boldsymbol{h}\left(\boldsymbol{m}_{\iTime \lvert \iTime-1}\right)$;  \\
   $\bvarytpred = \Jac \bvartpred\Jac^\tra +\bR$;  \\
$\boldsymbol{K}_{\iTime}=\bvartpred\, \Jac^\tra\, \bvarytpred^{-1}$;  \\
$\bPSAt\left( \Model \lvert \byotprev \right) =\bvarytpred^{-1}$\\
\For{$\jAgent=1, \ldots, \NAgents$}{
$\cdf_{\RVhbyt}([\byt]_{\iData})= \frac{1}{2} \operatorname{erfc}\left( [\boldsymbol{e}_{\iTime}]_{\jAgent} / \sqrt{2\,[\bvarytpred}]_{\jAgent, \jAgent} \right)$; \\
 $d_{\iKS, \iData, \iTime}= \max\left\{\cdf_{\RVhbyt}([\byt]_{\iData}),\, 1-\cdf_{\RVhbyt}([\byt]_{\iData})  \right\}$; \\
 \noindent Compute the p-values through a one-sample \ac{KS} test under the null hypothesis that 
$d_{\iKS, \iData, \iTime} \sim \mathcal{U}([0.5, 1])$, as expected when the predictive model is correct. 
}
\For{$\jAgent=1, \ldots, \NAgents$}{
$\pdf_{\RVy}(\y_{\jAgent})= \mathcal{N}([\boldsymbol{y}_{\iTime}]_{\jAgent};[\boldsymbol{h}\left(\boldsymbol{x}_{\iTime }\right)]_{\jAgent},[\bR]_{\jAgent, \jAgent}) $; \\
$\pdf_{\RVhy} (\y_{\jAgent}) = \mathcal{N}([\boldsymbol{y}_{\iTime}]_{\jAgent};[\boldsymbol{h}\left(\boldsymbol{m}_{\iTime \lvert \iTime-1}\right)]_{\jAgent},[\bvarytpred]_{\jAgent, \jAgent})$; \\
}
$\dKLt=
\sum_{\jAgent} \pdf_{\RVy}(\y_{\jAgent}) \log\,\left(\pdf_\RVy (\y_{\jAgent}) / \pdf_{\RVhy} (\y_{\jAgent}) \right)$; \\
$\boldsymbol{m}_{\iTime \lvert \iTime}=\boldsymbol{m}_{\iTime \lvert \iTime-1} + \boldsymbol{K}_{\iTime}\,\boldsymbol{e}_{\iTime}$; \\
$\boldsymbol{P}_{\iTime \lvert \iTime}=\boldsymbol{P}_{\iTime \lvert \iTime-1}-\boldsymbol{K}_{\iTime}\,\bvarytpred\,\boldsymbol{K}_{\iTime}^\tra$; 
}
\caption{EKF with Proficiency Evaluation } \label{alg:alg1}
\end{algorithm}

\paragraph{The Kolmogorov-Smirnov (KS) Statistics}
The (two-sample) \ac{KS} statistics works differently from the \ac{MPB}, since it compares two \acp{CDF}, namely, $\cdf_{\RVy}$ and $\cdf_{\RVhy}$, or their empirical versions, namely $\approxcdf_{\RVy}^{[\NSampleL]}$ and $\approxcdf_{\RVhy}^{[\NSampleK]}$, where $\NSampleL$ and $\NSampleK$ are the number of collected samples.
The two \acp{CDF} are computed from two sets of data: $\approxcdf_{\RVy}^{[\NSampleL]}(\y)$ considers $\NSampleL$ samples of the actual measurement $\y \in \left\{ \y^{(1)}, \y^{(2)},\, \ldots, \, \y^{(\NSampleL)} \right\}$, while $\approxcdf_{\RVhy}^{[\NSampleK]}(\y)$ refers to their predictions $\y \in \left\{\hy^{(1)}, \hy^{(2)},\, \ldots, \, \hy^{(\NSampleK)} \right\}$, drawn from the corresponding predictive distribution of measurements. More specifically, the \ac{KS} statistic is defined as \cite{djuric2010assessment}
\begin{equation}\label{eq:dsLK}
    {\RVdKS^{[\NSampleL,\NSampleK]}} \triangleq \sup\limits_{\y}\, \left\lvert \approxcdf_{\RVy}^{[\NSampleL]}(\y) -\approxcdf_{\RVhy}^{[\NSampleK]}(\y) \right\rvert.
\end{equation}
We can now define the following hypothesis test 
\begin{align}
        &\Hnull: \cdf_{\RVy}^{[\NSampleL]}(\y; \CModel)=\cdf_{\RVhy}^{[\NSampleK]}(\y; \Model),  \nonumber\\
        &\Ha: \cdf_{\RVy}^{[\NSampleL]}(\y; \CModel)\neq \cdf_{\RVhy}^{[\NSampleK]}(\y; \Model), \label{eq:KS_test}
\end{align}
where the hypothesis $\Hnull$ is rejected (i.e., we decide $\mathcal{D}_1$ which means that  $\CModel \neq \Model$), at a level $\alpha$ if
%
$\sqrt{\frac{\NSampleL\,\NSampleK}{\NSampleL+\NSampleK}}\, d_{\mathrm{KS}}^{[\NSampleL,\NSampleK]} \geq \gamma_{\alpha}$
%
where $\gamma_{\alpha}$ is a threshold \cite{djuric2010assessment, rohatgi2015introduction}. 

If we consider the collection of only a single sample of the actual measurement (i.e., $\NSampleL=1$) and that $\NSampleK>1$ predictions are drawn from the predictive distribution, \eqref{eq:dsLK} becomes
\begin{align}\label{eq:ds}
    \! \RVdKS^{[1,\NSampleK]}=\RVdKS 
    &= \max\left\{\approxcdf_{\RVhy}^{[\NSampleK]}(y^{(1)}),\, 1-\approxcdf_{\RVhy}^{[\NSampleK]}(y^{(1)}) \right\},
\end{align}
with the support of $\RVdKS$ given by $0.5 \leq \dKS \leq 1$. Moreover, if $\y^{(1)}$ and $\hy^{(k)},$ with $ \iSampleK=1, \ldots, \NSampleK$, are i.i.d. samples from the same distribution, then $\RVdKS^{[1,\NSampleK]}$ is a discrete uniform \ac{RV}, as shown in \cite[Prop. 2]{djuric2010assessment}. 
\begin{figure}[t!]
\centering
\input{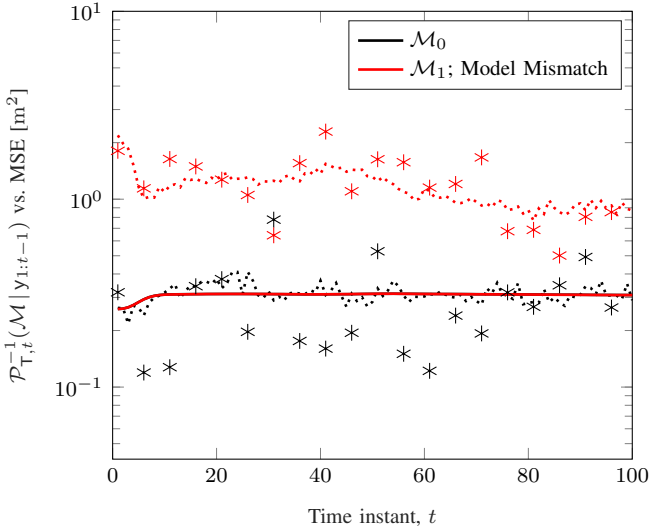}
    \caption{Inverse of the team \ac{MPB} according to the selected models and as a function of time. Markers plot the error ${e}_t^2\left(\Model\right)$ obtained from a single simulation trial, and dotted lines depict the \ac{MSE} averaged over $100$ trials. 
    }
    \label{fig:invPSA_MA_B}
\end{figure}
\begin{figure*}[t!]
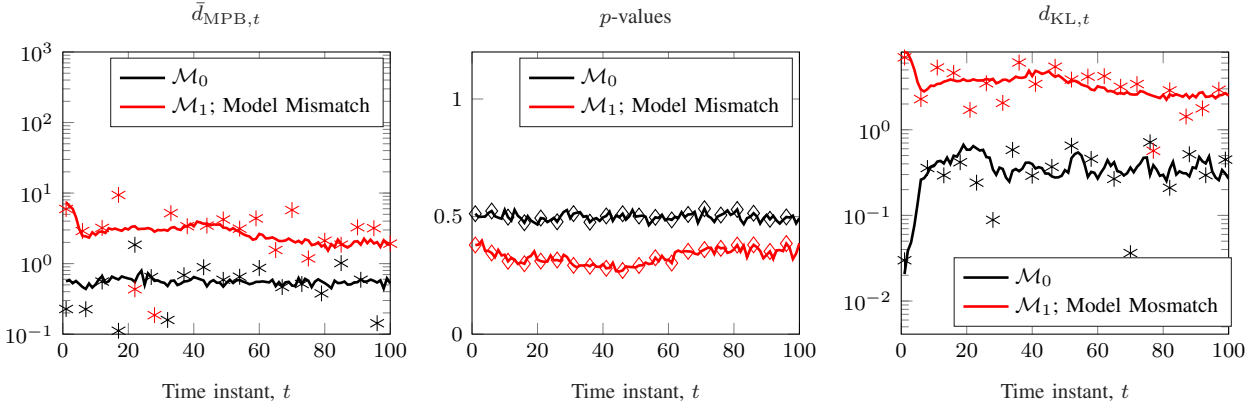

\input{dPSA_MM_B_MA}
\input{pvalues_MM_B_MA}
\input{dKL_MM_B_MA}
\caption{Left: $\ndPSAt$ as a function of $t$ and $\Model$.  Middle: Sequence of averaged $p$-values for the \ac{KS}. Right: \ac{KL} divergence. Continuous lines represent the distance averaged over $\NMC=100$, whereas markers refer to a single simulation trial. 
}
\label{fig:dPSApVdKL_MA_B}
\end{figure*}
Given the model assessment problem in \eqref{eq:KS_test}, it is possible to assess the proficiency of a model using the \ac{KS} statistics by running a standard one-sample \ac{KS} test resulting in a sequence of $p$-values. Each $p$-value represents the highest level of Type I error probability acceptable under which the null hypothesis remains accepted (assuming it is true).
In this sense, the $p$-value can be interpreted as a metric of proficiency, as it indicates how small the significance level $\alpha$ must be to reject the null hypothesis.
Small $p$-values indicate a weak alignment between the data and the model. 

Starting from \eqref{eq:ds}, it is straightforward to extend the \ac{KS} statistics to a team by writing the \ac{ECDF} of the joint likelihood distribution \cite{rohatgi2015introduction, djuric2010assessment}. Alternatively, one can evaluate the distance for each element of the observation vector as
\begin{equation}\label{eq:dKS_L}
    d_{\iKS, \iData}= \max\left\{\cdf_{\RVhby}([\by]_{\iData}),\, 1-\cdf_{\RVhby}([\by]_{\iData})  \right\},
\end{equation}
with $\iData=1, \ldots, \NAgents$ being the index of the observations collected by each agent, and where $\cdf_{\RVhby}([\by]_{\iData})$ is the measurement predictive \ac{CDF} evaluated for the $\iData$-th entry of $\by$.  

\paragraph{The Kullback-Leibler Divergence}

The \ac{KL} divergence is a statistical measure that quantifies the proximity of the \ac{PDF} describing the actual measurement, namely $\pdf_{\RVby}(\cdot)$, to a candidate \ac{PDF}, namely $\pdf_{\RVhby}(\cdot)$ \cite{cover1999elements}. Since $\pdf_{\RVby}(\cdot)$ depends on the true state $\RVbxt$, which is not available in practice, the \ac{KL} divergence can only serve as a theoretical benchmark for assessing model performance. One method to evaluate a candidate distribution, based on the model choice $\Model$, is to consider the value of $\dKL$ since very low values indicate good agreement between the two models. 

\section{Case Study: Proficiency Evaluation for Target Tracking}
\label{sec: casestudy_vector}

We now consider a non-linear Gaussian vector \ac{SSM} where the team of \acp{AA}, placed in known positions $\pos_{\iAgent,\iTime}$, with $\iAgent=1, 2, \ldots, \NAgents$, at time instant $\iTime$, cooperates to track the state of a moving target \cite{guerra2020dynamic}. We define the following \ac{SSM}:
\begin{align}\label{eq:SSM_MA}
    \begin{cases}
        \RVbxt = \bA\, \RVbxtprev + \RVbut, \\
        \RVyit = \hit (\RVbxt)  + \RVvit, \quad  \RVyit \in \RVbyt, 
    \end{cases}
\end{align}
where $\RVbxt=[\RVpost,\, \RVspeedt]^\tra \in \mathbb{R}^{\NState \times 1}$ is the state of the target (i.e., its position and velocity), $\NState$ is the size of the state (e.g., for 2D target tracking, it is $\NState=4$), $\bA \in \mathbb{R}^{\NState \times \NState}$ and $\RVbut \sim \Normal(\boldsymbol{0}, \bQ)$ are the state transition matrix and the state noise process with covariance matrix $\bQ$, respectively, set as in \cite{sarkka2023bayesian}
\begin{align}\label{eq:transition_parameters}
\bA= \left[\begin{array}{cc}
     \mathbf{I}_2 & \dt \mathbf{I}_2  \\
     \mathbf{0}_2 & \mathbf{I}_2
\end{array} \right], \quad \bQ=\left[\begin{array}{cc}
     \frac{\dt^3}{3}\, \bQQ &  \frac{\dt^2}{2}\,\bQQ  \\
     \frac{\dt^2}{2}\,\bQQ & \dt \bQQ 
\end{array} \right],
\end{align}
with $\dt$ being the sampling interval and $\bQQ=\diag([q_x,\, q_y])$ containing the variances of the changes in accelerations. In the observation equation,  $\hit (\RVbxt)$ is the observation function, and $\RVvit \sim \Normal(0, \sigma^2_{v,i})$ is the measurement noise process.

By applying the traditional \ac{EKF} for tracking the state, the predictive distribution is given by
$\pdf_{\RVbyt \lvert  \RVbyotprev}(\byt \lvert \byotprev; \Model) \! =\!\Normal\!\left(\RVbyt;\, \bht(\bmtpred),  \bvarytpred \right),$ where $\bht=[{h}_{1,\iTime}, \ldots, {h}_{\iAgent,\iTime}, \ldots, {h}_{\NAgents,\iTime} ]^\tra$, $\bmtpred$ is the predicted state obtained using the state transition equation. The innovation covariance matrix is given by
\begin{align}\label{eq:S}
\bvarytpred = \Jac(\bmtpred) \bvartpred\Jac^\tra(\bmtpred) +\bR,
\end{align}
where $\Jac(\bmtpred)$ is the Jacobian matrix evaluated at the predicted state (i.e., at $\bmtpred$), $\bvartpred$ is the predicted state covariance matrix, and $\bR=\diag(\sigma^2_{v, 1}, \ldots, \sigma^2_{v, \iAgent}, \ldots, \sigma^2_{v, \NAgents})$ is a diagonal matrix containing the noise observation variances of each agent.
For this \ac{SSM}, the \ac{MPB} is
\begin{align}\label{eq:proficiencySSM}
\bPSAt\left( \Model \lvert \byotprev \right) \approx\bvarytpred^{-1}.
\end{align}
Note that in linear/Gaussian \ac{SSM}, the covariance 
$\bvarytpred$ coincides with the true innovations covariance; thus, 
$\bvarytpred^{-1}$ represents the exact \ac{MPB}. 
Instead, in the nonlinear case, $\bvarytpred$ results from a local 
linearization and only approximates the true predictive covariance, 
so $\bvarytpred^{-1}$ should be regarded as an approximation to \ac{MPB}.
Considering \eqref{eq:S} and \eqref{eq:proficiencySSM}, we can observe that the team proficiency depends on two key components. The first term accounts for the system geometry through the Jacobian matrix $\Jac(\cdot)$, the uncertainty of the state estimate, and the transition model through the predicted state covariance matrix $\bvartpred$, while the second term depends on the measurement covariance matrix.  

\section{Simulation Results} 
\label{sec:numresults}
We consider four collaborative agents located at fixed positions, i.e., $\pos_1=[-10,\, 0]^\tra$, $\pos_2=[-10,\, 30]^\tra$, $\pos_3=[20,\, 0]^\tra$, $\pos_4=[20,\, 30]^\tra$, in meters. 
Each agent collects a range measurement at time instant $\iTime$, so that
\begin{align}\label{eq:meas}
&\hit (\RVbxt)= \mathsf{d}_{\iAgent, \iTime}(\posi, \RVpost)+ \mathsf{b}_{\iAgent, \iTime}(\posi, \RVpost), 
\end{align}
where $\mathsf{d}_{\iAgent, \iTime}(\posi, \RVpost)=\lVert \posi - \RVpost \rVert$ is the agent-target distance and $\mathsf{b}_{\iAgent, \iTime}(\posi, \RVpost)$ is the ranging bias being $0$ if the measurement link between the agent and the target is in \ac{LOS}, and $\mathsf{b}_{\iAgent, \iTime} >0$, otherwise.
The ranging noise is given by $\RVvit \sim \Normal(0, \varvit)$ with $\varvit=\varv$, $\forall \iAgent=1, 2, \ldots, \NAgents$ and $\forall \iTime=1,2,  \ldots, \T$, which gives $\bR=\varv\, \boldsymbol{I}_{\NAgents}$.

The true target state is given by $\bxt=[\post^\tra, \, \speedt^\tra]^\tra$ initialized with $\bx_{0}=[\pos_{0}^\tra, \, \speed_{0}^\tra]^\tra=[\spos_{x,0},\, \spos_{y, 0},\, \dot{\spos}_{x,0},\, \dot{\spos}_{y, 0} ]^\tra=[\boldsymbol{0}_2^\tra,\, [0.2,\, 0.4]^\tra ]^\tra$.
For generating its trajectory, we consider \eqref{eq:transition_parameters} with $\dt=1\, \mathrm{s}$, $q_x=\left({\dot{\spos}_{x,0}}/{\beta_0}\right)^{2} \,\dt^{-1}\, \mathrm{m}^2/\mathrm{s}^3$, $q_y=\left({\dot{\spos}_{y,0}^2}/{\beta_0}\right)^{2} \, \dt^{-1}\, \mathrm{m}^2/\mathrm{s}^3$, with $\beta_0=10$, and we work with  $\T\!\!=\!\!100$ instants and $\NMC\!\!=\!\!100$. 
We use an \ac{EKF} to estimate the target's trajectory, as described in Alg.~\ref{alg:alg1}. We set $\bm_{0 \lvert 0}=\bxo$ and $\bP_{0 \lvert 0}= \diag([0.1^2,\, 0.1^2,\, \left({\dot{\spos}_{x,0}}/{100}\right)^2,\, \left({\dot{\spos}_{y,0}}/{100}\right)^2 ]^\tra )$.

We consider a mismatch in the observation model by varying the knowledge about the NLOS ranging biases, which typically represent the main sources of error. 
More specifically, we consider a true generative model given by $\Model_0=\{\Model_{0,x},\, \Model_{0,y}  \}$, $\forall \iTime$, where $\Model_{0,x}$ follows the transition model in \eqref{eq:transition_parameters},  and  $\Model_{0,y}: \btheta_{0_y} = \left\{\sigma_v, \left\{ b_{\iAgent, \iTime} \lvert \iAgent=1, 2. \ldots, \NAgents \right\} \right\} = \left\{0.5,\, 0,\,    2.7,\,         0,\,    0.07 \right\} \mathrm{m}$. This corresponds to a scenario with the agents in $\pos_2$ and $\pos_4$ in \ac{NLOS}. 
For state estimation, we adopt the following two models: $\Model_0:\{ \btheta_{0,x}, \btheta_{0,y} \}$, and  
$\Model_1=\{\Model_{0,x},\, \Model_{1,y}  \}:\{ \btheta_{0,x}, \btheta_{1,y} \},$ with $\btheta_{1,y}=\left\{0.5,\, 0,\, 0,\, 0,\, 0\right\}$. The first model assumes a perfect \ac{CSI} whereas the second one wrongly assumes that the agents are always in \ac{LOS}  despite the true propagation conditions.  
Figure~\ref{fig:invPSA_MA_B} reports: (i) the inverse of the \ac{MPB} according to the selected models (solid lines); the error ${e}_t^2\left(\Model\right)$ obtained from a single simulation trial (markers); and the \ac{MSE} averaged over $100$ trials (dotted lines). The inverse of the \ac{MPB} is the same for the two models, but the agreement between its inverse and the \ac{MSE} holds only when the algorithm has perfect knowledge about the \ac{NLOS} ranging biases. 
The corresponding metrics in terms of normalized \ac{MPB} distance, $p$-values, and \ac{KL} divergence are reported in Fig.~\ref{fig:dPSApVdKL_MA_B}. As expected, the use of the correct model $\CModel$ is associated with lower values of $\ndPSAt$ and $\dKLt$, and the $p$-values distribute around $0.5$.
The results suggest that, even without knowledge of the true model, both the \ac{MPB} and the \ac{KS} provide a robust assessment of model proficiency.

\section{Conclusions}
\label{sec:conclusions}
This letter investigated the problem of proficiency assessment for a centralized team of \acp{AA}. 
We proposed a comparative analysis based on three metrics: the \ac{MPB}, the \ac{KS} statistic, and the \ac{KL} divergence. 
While the \ac{KL} divergence was adopted as a ground-truth reference, the \ac{MPB} and \ac{KS} metrics provide practical and computationally efficient alternatives for \emph{in situ} assessment. 
Results in a target tracking scenario demonstrated that both \ac{MPB} and \ac{KS}-based indicators successfully capture model mismatch and align well with the \ac{KL}-divergence benchmark, even without access to the ground truth state. 
Future work will extend this framework to distributed teams.

\clearpage
\newpage

\bibliographystyle{IEEEtran}

\end{document}